\def\mycmd{2}
\if\mycmd1
\documentclass[11pt,onecolumn, draftcls]{IEEEtran}
\else 
\documentclass[10pt,a4paper,twoside,journal]{IEEEtran}
\fi
\usepackage{tabularx}

\usepackage{multicol}
\usepackage{multirow}
\usepackage{float}
\usepackage{mathtools}
\usepackage{amsmath}
\usepackage{amsthm}
\usepackage{amssymb}
\usepackage{graphicx}
\usepackage{transparent}
\usepackage{wasysym}
\usepackage{setspace}
\usepackage{color}
\usepackage{bm}
\usepackage{cases}
\usepackage{booktabs}
\usepackage{adjustbox}
\usepackage{subfigure}
\usepackage{glossaries}
\usepackage{kotex}

\if\mycmd1
\usepackage[top=1.6cm, left=1.6cm, bottom=1.6cm, right=1.6cm]{geometry}
\else
\usepackage[top=1.4cm, left=1.4cm, bottom=1.4cm, right=1.4cm]{geometry}
\fi
\usepackage{tikz}
\newcommand*\circled[1]{\tikz[baseline=(char.base)]{
            \node[shape=circle,draw,inner sep=.3pt] (char) {#1};}}
\usepackage{amsfonts}
\usepackage{makecell}
\usepackage{pbox}
\usepackage{epsfig}

\makeatletter

\floatstyle{ruled}
\newfloat{algorithm}{tbp}{loa}
\providecommand{\algorithmname}{Algorithm}
\floatname{algorithm}{\protect\algorithmname}

\theoremstyle{plain}

\theoremstyle{plain}

\theoremstyle{plain}

\usepackage{cite}
\usepackage{stfloats}
\usepackage{graphicx}
\usepackage{array}
\usepackage{enumerate}
\usepackage{enumitem}
\setlist[itemize]{leftmargin=*}
\setlist[enumerate]{leftmargin=*, label=\arabic*)}
\usepackage{graphicx}
\usepackage{times}
\usepackage{algorithm}
\usepackage{algpseudocode}
\theoremstyle{remark}

\newtheorem*{remark}{Remark}

\makeatother

\algrenewcommand\algorithmicindent{1.0em}%
\providecommand{\lemmaname}{Lemma}
\providecommand{\propositionname}{Proposition}

\providecommand{\theoremname}{Theorem}
\providecommand{\theoremname}{Definition}
\newcommand{\rom}[1]{\uppercase\expandafter{\romannumeral #1\relax}}

\algnewcommand{\IIf}[1]{\State\algorithmicif\ #1\ \algorithmicthen}
\algnewcommand{\ElseIIf}[1]{\algorithmicelse\ #1} 
\algnewcommand{\EndIIf}{\unskip\ \algorithmicend\ \algorithmicif}

\newcounter{problem}
\newcounter{save@equation}
\newcounter{save@problem}


\numberwithin{save@problem}{subsection}
\numberwithin{save@equation}{subsection}

\begin{document}
\title{Adaptive Resource Allocation Optimization Using Large Language Models in Dynamic Wireless Environments}
\author{Hyeonho Noh,~\IEEEmembership{Member,~IEEE}, Byonghyo Shim,~\IEEEmembership{Fellow,~IEEE}, and Hyun Jong Yang,~\IEEEmembership{Member,~IEEE}
\thanks{Hyeonho Noh, Byonghyo Shim, and Hyun Jong Yang are with Department of Electrical and Computer Engineering, Seoul National University, Seoul, 08826, Korea (e-mail: hyeonho@postech.ac.kr, bshim@islab.snu.ac.kr, hjyang@snu.ac.kr).
Hyun Jong Yang is the corresponding author.}
}
\maketitle
\begin{abstract}\label{abstract}
Deep learning (DL) has made notable progress in addressing complex radio access network control challenges that conventional analytic methods have struggled to solve. 
However, DL has shown limitations in solving constrained NP-hard problems often encountered in network optimization, such as those involving quality of service (QoS) or discrete variables like user indices. Current solutions rely on domain-specific architectures or heuristic techniques, and a general DL approach for constrained optimization remains undeveloped. Moreover, even minor changes in communication objectives demand time-consuming retraining, limiting their adaptability to dynamic environments where task objectives, constraints, environmental factors, and communication scenarios frequently change. To address these challenges, we propose a large language model for resource allocation optimizer (LLM-RAO), a novel approach that harnesses the capabilities of LLMs to address the complex resource allocation problem while adhering to QoS constraints. By employing a prompt-based tuning strategy to flexibly convey ever-changing task descriptions and requirements to the LLM, LLM-RAO demonstrates robust performance and seamless adaptability in dynamic environments without requiring extensive retraining. Simulation results reveal that LLM-RAO achieves up to a $40$\% performance enhancement compared to conventional DL methods and up to an $80$\% improvement over analytical approaches. Moreover, in scenarios with fluctuating communication objectives, LLM-RAO attains up to $2.9$ times the performance of traditional DL-based networks.

\end{abstract}


\section{Introduction}
\label{sec:introduction}

Over the past few years, there has been growing interest in using deep learning (DL) techniques in handling complex radio access network control problems, such as resource allocation, power control, and interference management \cite{Yuanming23_WC, Alessio19_TCOM}. While we are witnessing remarkable advantages of DL-based systems, there are still fundamental limitations in their capability to tackle complex optimization problems within network management effectively. One main reason is that while the DL approach is effective in unconstrained problem settings, it might not work well in solving NP-hard problems with stringent constraints (e.g. quality of service (QoS)) or discrete variables (e.g. user indices) \cite{Jose22_NIPS, Donti21_ICLR}. Since most network optimization problems are constrained optimization problems, they should be handled through carefully tailored DL techniques developed by domain experts or managed using heuristic techniques such as penalty regularization \cite{Jose22_NIPS} or solution projection \cite{Donti21_ICLR}. 
Thus, a universally applicable DL methodology to solve the constrained optimization problems has yet to be established.
Another important reason is that DL requires a time-consuming retraining process, even with slight changes in the objective function, constraints, or communication environment, to adapt to the modified scenario.

Future communication systems will evolve beyond merely transmitting data to actively support specific computations and tasks \cite{Yalin23_WC, Yuanming23_WC, Deniz23_JSAC}. While conventional systems have focused on improving data transmission rates and latency, next-generation communication systems will prioritize ensuring the QoS needed for communication users to make accurate inferences, decisions, and actions at the right time to achieve successful task completion. 
Furthermore, task-related QoS requirements, communication variables (e.g., number of users), and communication objectives (e.g., data rate or proportional fairness) will dynamically shift in response to changing task goals.
This signifies a departure from conventional fixed DL-based approaches tailored for unchanged environments, moving toward a new paradigm-one that can flexibly adapt to changing environments and objectives, seamlessly adjusting the network to sustain strong and resilient performance.

To deal with the deficiencies of the existing fixed DL-based approaches, the use of large language models (LLMs) has recently garnered considerable attention recently. They disclose the potential of LLMs in solving communication system optimization problems such as power allocation \cite{Jiawei24_JCIN, Jingwen24_arxiv}. However, existing works have addressed relatively simple system models, tackled less complex problems, and overlooked the LLMs' adaptability to ever-changing environments and task objectives, their most significant advantage.

This paper tackles highly complex system optimization problems that maximize objective functions while adhering to QoS constraints and queuing model under dynamically changing environments, such as maximizing the proportional fairness of wireless cellular networks while considering the number of users' backlogged packets. In our work, we propose a prompt-based tuning of LLM for the resource allocation optimizer (LLM-RAO). This approach addresses optimization problems involving not only resource allocation but also the joint optimization of orthogonal frequency division multiple access (OFDMA), mode selection between single-user multi-input-multi-output (SU-MIMO) and multi-user MIMO (MU-MIMO), MU-MIMO user selection, and channel selectivity. 

Our contributions are as follows:
1) We propose a prompt-based LLM tuning method to fine-tune LLM-RAO and address highly complex constrained communication optimization problems. To effectively utilize an off-the-shelf LLM, which is not inherently tailored to tackle mathematical optimization problems, we implement a closed-loop prompting method that autonomously generates prompts, performs inference, and iteratively refines the RA solution based on the inference history. The prompt-based LLM tuning also enables LLM-RAO to flexibly adjust to changing task objectives (e.g., data rate or proportional fairness), constraints (e.g., minimum rate requirements), environmental factors (e.g., the number of users), and communication scenarios (e.g., buffered queue status). 2) To evaluate the problem-solving capability and adaptability of LLM-RAO in various wireless scenarios where objective functions and constraints are shifting dynamically, we conduct experiments on the resource allocation optimization problem for uplink communication in IEEE 802.11ax. The simulation results show that LLM-RAO achieves up to a $40 \%$ performance improvement over a conventional deep reinforcement learning (DRL)-based approach and up to an $80 \%$ improvement over an analytical method. Additionally, LLM-RAO achieves up to 2.9 times higher performance compared to a fixed DL-based network in environments with changing communication objectives.

\section{System Model and Problem Formulation}
\label{sec:system_model}

\begin{figure}[t]
    \centering
    \includegraphics[draft=false, width= \if 1\mycmd 0.7 \else 0.99 \fi \columnwidth]{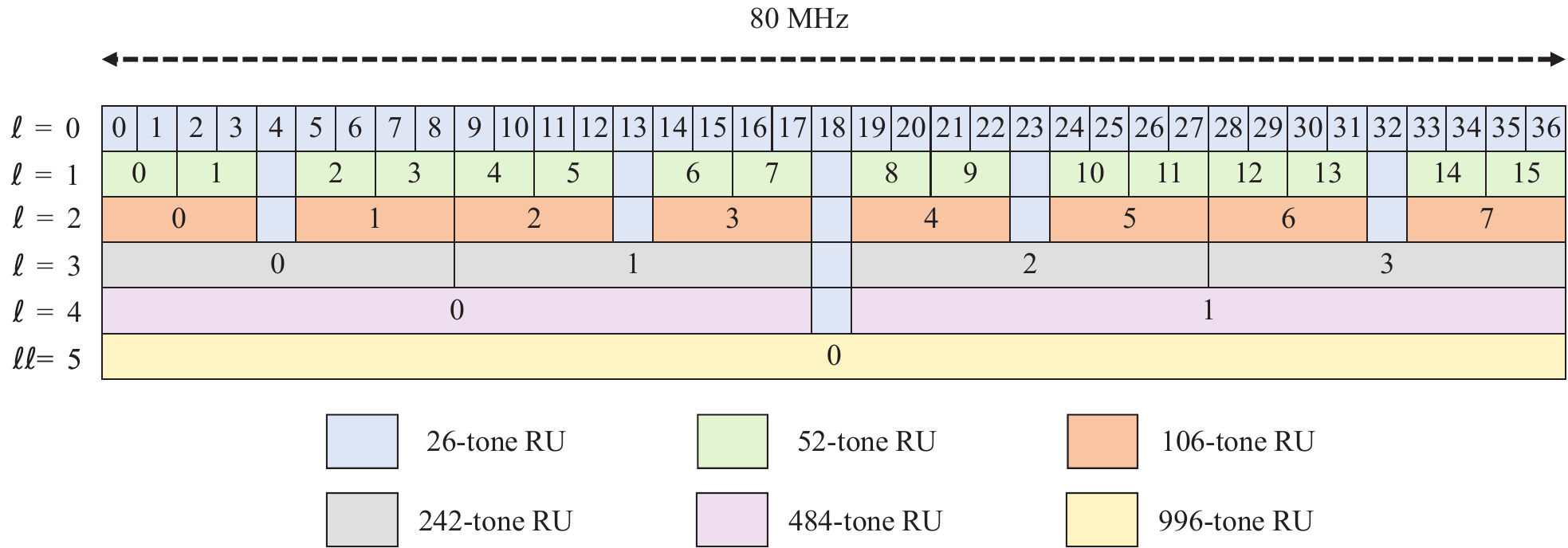}
    \vspace{-12pt}
    \caption{OFDMA RBs in a 20 MHz channel.}
    \label{fig:RB_table}
    \vspace{-12pt}
\end{figure}

In the context of IEEE 802.11ax uplink, we consider a basic service set where an access point (AP) communicates with $K$ users. The AP is equipped with $N$ antennas, while each user employs a single antenna. The AP allocates the users to $\text{RB}(l,i)$, which denotes the resource block (RB) at the $l$-th level and $i$-th index, as shown in Fig. \ref{fig:RB_table}. To represent the allocation $\text{RB}(l,i)$ to user $k$, a binary variable is defined as $x_{l,i}^{(k)}$ such that $x_{l,i}^{(k)} = 1$ if the $k$-th user is allocated to $\text{RB}(l,i)$, and $x_{l,i}^{(k)} = 0$ otherwise.

\subsection{OFDMA RB Allocation}
\label{subsec:ofdma_ru_allocation}

The IEEE 802.11ax WLAN standard \cite{IEEEstd21} has established the following constraints in RB allocation: 1) a single user or group of users cannot be allocated to multiple RBs; 2) the simultaneous allocation of multiple RBs on the same frequency band is forbidden; 3) MU-MIMO is available only in RBs with sizes greater than 52-tone. To represent the constraint 2, we define $\mathbf{e}_{l,i} \in \mathbb{Z}^{q \times 1}$ as an indicating vector, the $n$-th element of which is defined such that $[\mathbf{e}_{l,i}]_n \hspace{-2pt}=\hspace{-2pt}1$ if $\text{RB}_{l,i}$ overlaps with $\text{RB}_{0,n}$ in the spectrum domain, and $[\mathbf{e}_{l,i}]_n \hspace{-2pt}=\hspace{-1pt} 0$ otherwise, where $q$ is the number of 26-tone RBs, and $[\mathbf{x}]_n$ is the $n$-th entry of a vector $\mathbf{x}$.
Then, we can represent the constraints on the RA as follows:
\begin{subequations}\label{eq:constraints}
    \begin{align}
            & \text{Constraint 1: } \sum_{\forall (l,i)} x_{l,i}^{(k)} \leq 1, ~~~\forall k \in \mathcal{K}, \label{subeq:1a} \\
            & \text{Constraint 2: }\sum_{\forall(l,i)} \delta\left( \sum_{k \in \mathcal{K}} x_{l,i}^{(k)} \mathbf{e}_{l,i} \right) \preceq \mathbf{1}_q, \label{subeq:1b} \\
            & \text{Constraint 3: } \sum_{k \in \mathcal{K}} x_{l,i}^{(k)} \leq G(l), ~~~\forall (l, i), \label{subeq:1c}
    \end{align}
\end{subequations}
where {$\delta: \mathbb{Z}^{q \times 1} \rightarrow \mathbb{Z}^{q \times 1}$ }is the function defined such that $[\delta(\mathbf{x})]_n = \min\left\{ 
1, \left[\mathbf{x}\right]_n \right\}$ for a vector $\mathbf{x}$, $\mathbf{1}_q$ is the $q$-dimensional vector defined by $\mathbf{1}_q=[1, 1, \ldots, 1]^\text{T}$, $G$ is the function such that $G(l) = 1$ if $\text{RB}_{l,i}$ is either 26- or 52-tone RB, and $G(l) = N$ otherwise, and $\preceq$ is the element-wise inequality operator.

\subsection{Data Rate and Queue Model}
\label{subsec:throughput_model}

We denote $\mathbf{h}_{l,i}^{(k)} \in \mathbb{C}^{N \times 1}$ as the uplink communication channel between the AP and the $k$-th user on $\text{RB}_{l,i}$. 
The receive beamformer $\mathbf{w}_{l,i}^{(k)} \in \mathbb{C}^{N \times 1}$ is applied to receive the signal from the $k$-th user on $\text{RB}_{l,i}$.
Then, the signal-to-interference-plus-noise-ratio (SINR) for the $k$-th user on $\text{RB}_{l,i}$ yields $\Gamma_{l,i}^{(k)} = \frac{P_{l,i}^{(k)} \left| \left( \mathbf{w}_{l,i}^{(k)} \right)^\text{H} \mathbf{h}_{l,i}^{(k)} \right|^2_2 }{ \sum_{{m \in \mathcal{K} \backslash \{k\}}} P_{l,i}^{(m)} \left| \left( \mathbf{w}_{l,i}^{(k)} \right)^\text{H} \mathbf{h}_{l,i}^{(m)} \right|^2_2  + \sigma^2}$,
where $P_{l,i}^{(k)}$ is the transmit power of user $k$ on $\text{RB}_{l,i}$, $\sigma^2$ is the noise variance, and $\mathcal{K} = \{0,1, \ldots, K-1\}$. The theoretical data rate of user $k$ on $\text{RB}_{l,i}$ subsequently reads $r_{l,i}^{(k)} = B_{l,i} \log (1+\Gamma_{l,i}^{(k)})$ (bits/s), where $B_{l,i}$ is the bandwidth of $\text{RB}_{l,i}$.

We consider a practical queuing model where users have finite backlogged packets.
Let $Q^{(k)}$ be the number of backlogged packets (bits) for user $k$. 
Since it is not possible to transmit data more than the number of backlogged packets, the (effective) data rate considering the backlogged packets can be represented by $\hat{r}^{(k)} = \min \{\sum_{\forall (l,i)} x_{l,i}^{(k)} r_{l,i}^{(k)}, Q^{(k)} / T \}$ where $T$ is the data transmission time.

\subsection{Problem Formulation}
The RA optimization problem to maximize the utility function $u$ corresponding to data rate $\hat{r}^{(k)}$ with the RA constraints, queuing model, and QoS constraint is given by
\begin{subequations}
\label{P:opt}
\begin{align}
    & \max_{\mathbf{X}} & & \sum_k u^{(k)}, \\
    & \text{s.t.} & & \eqref{subeq:1a} \text{-} \eqref{subeq:1c}, \\
    & & & \sum_{\forall (l,i)} r_{l,i}^{(k)} \geq x_{l,i}^{(k)} R^{(k)},
\end{align}
\end{subequations}
where $R^{(k)}$ is the minimum rate of the $k$-th user.
We examine two distinct utility functions, each tailored to a specific optimization objective: 1) $\sum_{k\in\mathcal{K}} \hat{r}^{(k)}$ aimed at maximizing the data rate and 2) $\sum_{k\in\mathcal{K}} \log (1+\hat{r}^{(k)})$ designed to achieve proportionally fair data transmission across users. 

The RA optimization problem (\ref{P:opt}), given its nature as mixed-integer non-linear programming, can be classified as NP-hard. Notably, the RA optimization problem (\ref{P:opt}) incorporates aspects such as OFDMA, MIMO mode selection, MU-MIMO user selection, queuing model, and QoS constraint. These elements make it challenging for LLMs to solve the problem efficiently.

\remark If the number of backlogged packets $Q^{(k)}$ is infinite, the problem to maximize the data rate can be converted to the problem of maximizing $\sum_{k \in \mathcal{K}} x_{l,i}^{(k)} r_{l,i}^{(k)}$, which corresponds to typical sum-rate maximization. When it comes to proportional fairness, the objective function in (\ref{P:opt}) transforms into $\sum_{k \in \mathcal{K}} \log (1+x_{l,i}^{(k)} r_{l,i}^{(k)})$, a commonly used expression for proportional fairness.

\section{Proposed LLM-based Resource Allocation Optimizer}
In this section, we propose an LLM-based resource allocation optimizer, named LLM-RAO, of which the LLM is adopted to solve the RA optimization problem in \eqref{P:opt}.

\begin{figure*}[t]
    \centering
    \includegraphics[draft=false, width= \if 1\mycmd 1.0 \else 1.0 \fi \textwidth]{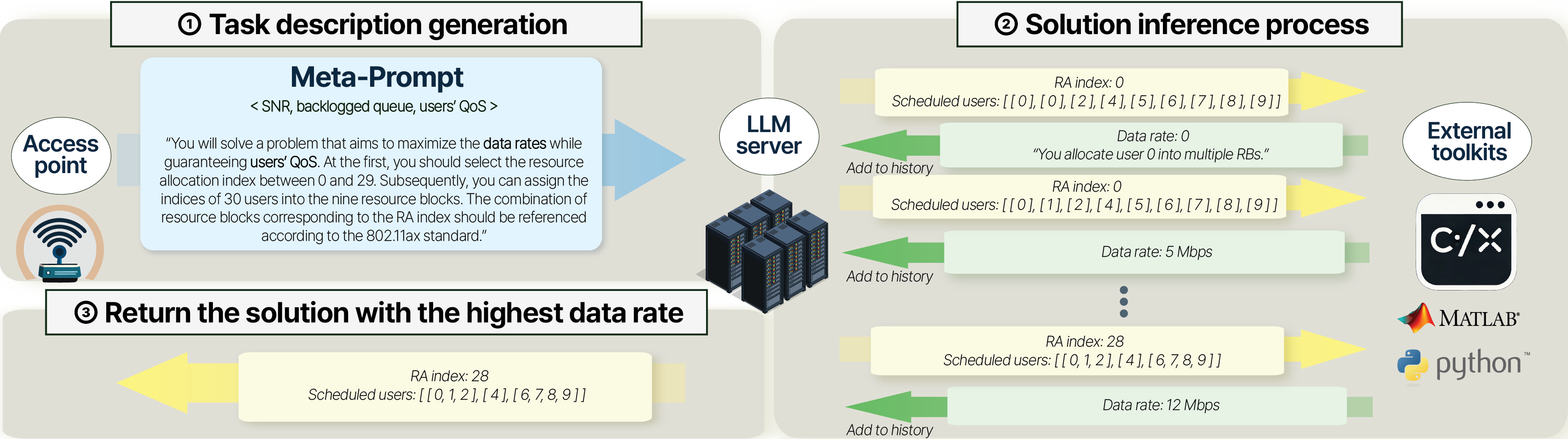}
    \caption{Overview of the proposed LLM-RAO. In the initial meta-prompt generation phase, the AP generates and transmits a meta-prompt including communication information of users and task description to the LLM server. During the solution inference process, the LLM server finds the optimal solution by iteratively generating solutions and receiving feedback on the score and constraint violations from the external toolkit. After completing the configured iterations, the LLM server delivers the final optimal solution to the AP.}
    \label{fig:system_model}
\end{figure*}

\subsection{Overview of LLM-RAO}

The RA optimization process can be summarized, as shown in Fig. \ref{fig:system_model}: 
The AP gathers the necessary communication specifications for RA, such as SINR, backlogged queue, and users' QoS, from the users. \circled{1} The AP then generates an initial meta-prompt that includes a description of the task environment and objective, encompassing the objective function, constraints, and the collected communication specifications. This generated meta-prompt acts as the guideline for the LLM server to understand the purpose and details of the communication task. The generated meta-prompt is then transmitted to an LLM server. 
\circled{2} Based on the received meta-prompt, the LLM server generates RA solutions aimed at maximizing the objective function. The RA results are evaluated by the external toolkits (e.g., MATLAB and Python), and the corresponding scores are provided to the LLM server via a prompt. When an RA result fails to meet any constraints, negative feedback is also relayed to the LLM, enabling further fine-tuning. The RA results and feedback are incorporated into the history of solutions and scores, and the LLM server generates new solutions by referring to the updated history. \circled{3} The optimization process finishes when the LLM server can no longer propose solutions with improved optimization scores or when the maximum number of optimization steps has been reached. Then, the LLM server returns a final RA solution.

\begin{figure}[t]
    \centering
    \includegraphics[draft=false, width= \if 1\mycmd 0.6 \else 1.0 \fi \columnwidth]{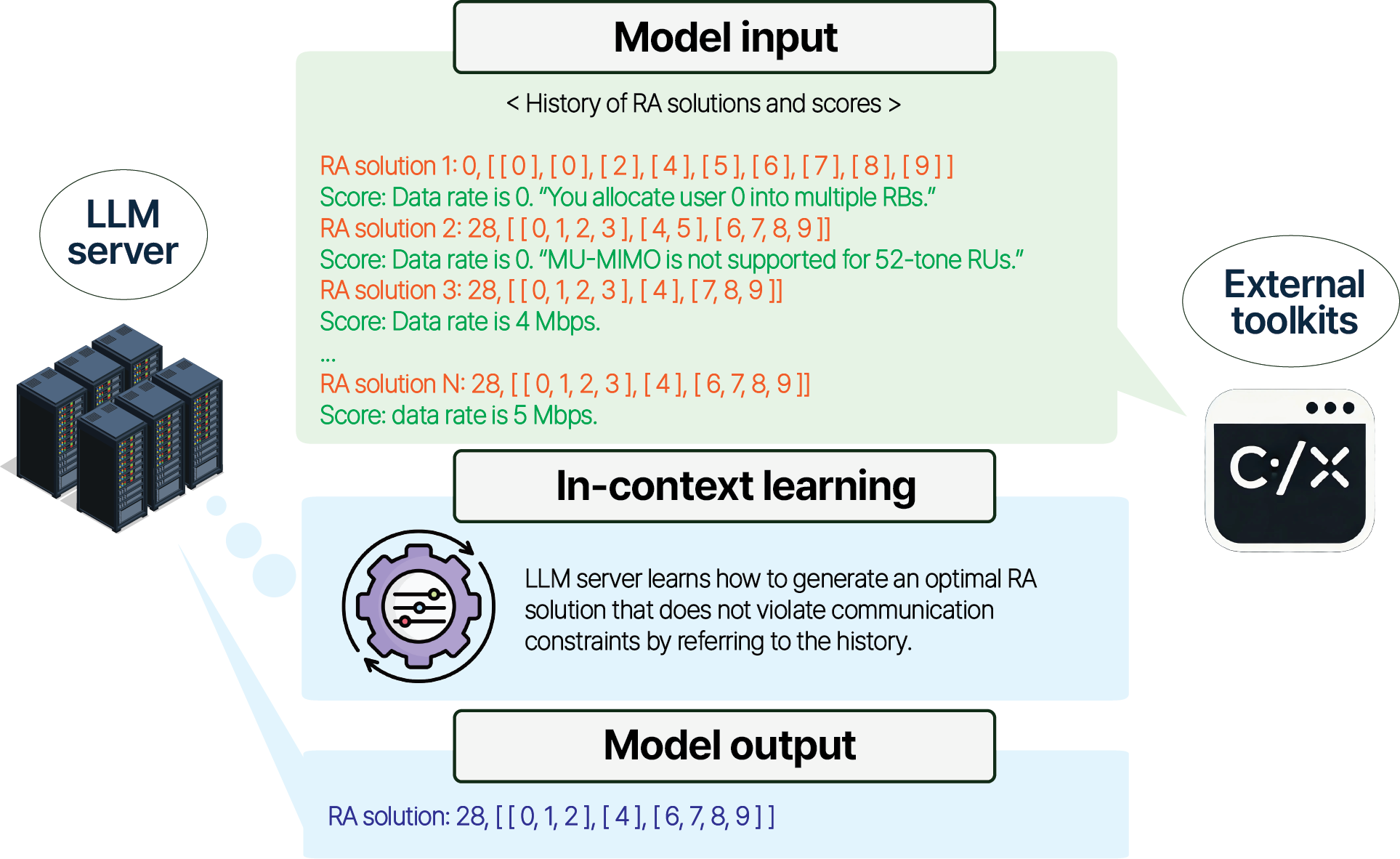}
    \vspace{-10pt}
    \caption{A detailed example of the inference process for solving the RA problem. The \textcolor{orange}{orange} text represents the solution generated by the LLM server, and the \textcolor{teal}{green} text describes the score for the generated solution as evaluated by external toolkits in the history of RA solutions and scores.}
    \label{fig:paragraph}
    \vspace{-10pt}
\end{figure}

\subsection{Task Description Generation}

As an initial step, the AP informs the LLM server regarding the task description and environment via meta-prompt. The meta-prompt serves as a critical guideline for the LLM server, ensuring a deep understanding of the task's goal and constraints. Therefore, it must be crafted with meticulous attention to detail, clearly defining the communication objective and system configuration from the outset. The process of generating an initial meta-prompt involves two essential components:

\textbf{Users' information:} 
The meta-prompt provides the LLM with key information about communication environment such as the users' channel, queue state, and QoS. 
Based on the information, the LLM understands the communication environment and solves the optimization problem to maximize the objective function.
It is worth noting that, unlike conventional DL-based methods, LLMs do not require human intervention in processing and applying information to solve optimization problems.
For example, when finding the optimal solution in \eqref{P:opt}, it is imperative to calculate and compare the data rates for all MU-MIMO user combinations. In the case of conventional DL-based methods, human manually derives the data rate formula based on the transceiver design and communication channel, and develops a DL model architecture suited to the optimization problem. This process not only demands significant humans' efforts but also results in a highly tailored model that lacks flexibility.
In contrast, when conveying channel information to LLMs, there is no need to manually calculate data rates, as LLMs can autonomously infer the data rates for all users by leveraging their trained internet-scale knowledge.
This approach substantially reduces the necessity for human involvement and improves the model's flexibility. 

\textbf{Communication objectives: }
To ensure that an LLM accurately solves an optimization problem, it is essential to have a meta-prompt that clearly defines the objective function, constraints, and optimization variables. 
Given the LLM's inherent limitation, known as ``hallucination," it might generate erroneous solutions that exceed permissible bounds. This issue is especially prevalent when the LLM is tasked with solving problems that were not included in its training data like complicated RA optimization problems. 
Therefore, it is essential for the AP to explicitly specify the range of optimization variables, such as ``user indices ($1$-$20$)," to prevent the LLM from generating invalid solutions, like referencing a non-existent user index such as 32.
By rigorously delineating these boundaries, the likelihood of the LLM producing out-of-scope solutions is minimized, thereby enhancing the precision and reliability of the optimization process.

In LLM-RAO, communication objectives can be easily adjusted by simply modifying the objective function, constraints, or system configuration within the meta-prompt. This flexibility allows for real-time adaptation to fluctuating conditions, such as varying communication objectives or network environments, without the help of complex reconfiguration or retraining processes. As a result, LLM-RAO can optimize resource allocation efficiently, maintaining high performance and meeting requirements even under dynamic environments.

\subsection{Inference Process}
Given that LLMs are language-based models, they are not inherently designed to directly address the RA optimization problem. To specialize the LLM for solving the RA problem, we employ a prompt-based fine-tuning technique called OPRO \cite{Yang24_ICLR}, which iteratively refines the solutions by iteratively generating solutions and providing feedback on the generated solutions accordingly. Thus, OPRO enables the LLM to internalize the core objectives and requirements of the RA problem, thereby facilitating the generation of an optimal RA solution.

Fig. \ref{fig:paragraph} illustrates a detailed example of this inference process. After the initial meta-prompt generation, a subset of the feedback, including solution, scores, and constraint satisfaction status, is sampled and incorporated into the history of RA solutions and scores. 
From the history of $<$solution, score$>$ pairs, the LLM server leverages in-context learning to discern complex patterns between solutions and scores, and iteratively refines its strategy to find the optimal solution. 
For example, the LLM server initially explores various solutions, some of which violate constraints, such as allocating the same user to multiple RBs or attempting unsupported MU-MIMO mode on 52-tone RBs. Based on the historical data, the LLM server recognizes these violations and adapts to the RA constraints. In addition, from the RA solutions $3$ and $N$, the LLM server can infer that allocating user $6$ is more advantageous for improving the data rate.
Through this process, the LLM server autonomously adjusts to the optimization problem while identifying critical factors including the constraints on RA. 



\begin{figure*}[t]
\centering
\subfigure[]{
\includegraphics[draft=false, width= \if 1\mycmd 0.4 \else 0.8 \fi \columnwidth]{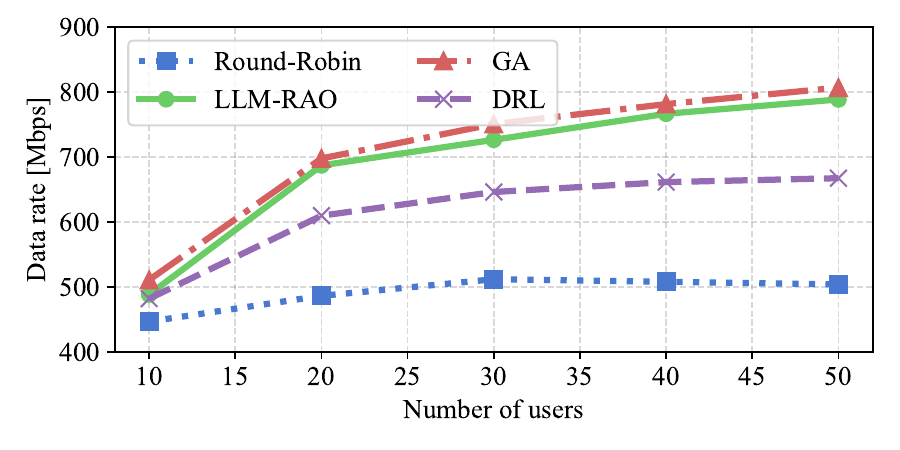}
\label{fig:sim1}
}
\subfigure[]{
\includegraphics[draft=false, width= \if 1\mycmd 0.4 \else 0.8 \fi \columnwidth]{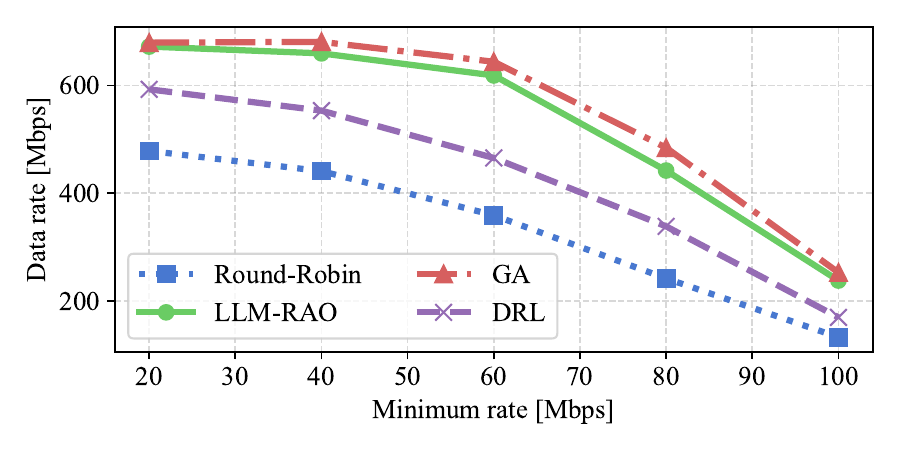}
\label{fig:sim2}
}
\subfigure[]{
\includegraphics[draft=false, width= \if 1\mycmd 0.4 \else 0.8 \fi \columnwidth]{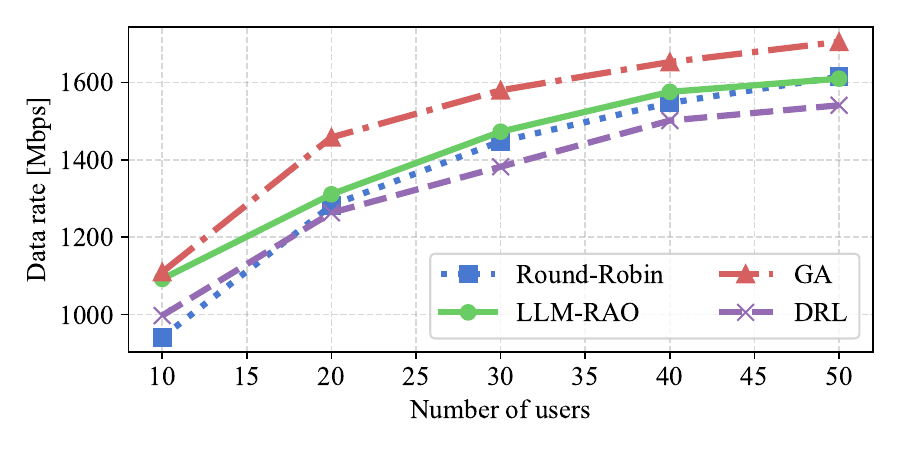}
\label{fig:sim4}
}
\subfigure[]{
\includegraphics[draft=false, width= \if 1\mycmd 0.4 \else 0.8 \fi \columnwidth]{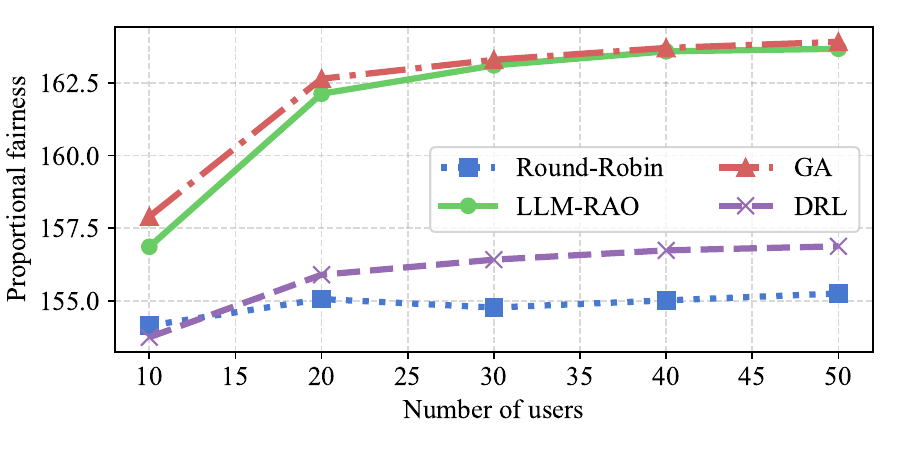}
\label{fig:sim3}
}
\vspace{-8pt}
\caption{Performance analysis of LLM-RAO and baseline methods in terms of data rate and proportional fairness under different scenarios. (a) scenario 1. (b) scenario 2. (c) scenario 3. (d) scenario 4.}
\label{fig:sim1234}
\vspace{-8pt}
\end{figure*}

\section{Simulation Results}
We have conducted simulations to evaluate the performance of the proposed LLM-RAO and compare it against baseline approaches. The simulation setup models an 802.11ax uplink system where an AP is equipped with $N = 4$ antennas. The system operates over a bandwidth of $20~\text{MHz}$, corresponding to 9 RBs with a transmission time of $T=4.848~\text{ms}$. For the simulations, we have developed a MATLAB simulator following the IEEE 802.11 standard channel model specifications \cite{80211_channel}. The users are randomly positioned within a distance range of $10$ meters to $100$ meters from the AP. The number of backlogged packets for each user is generated uniformly within the range of 0 to 50.
We have utilized OpenAI's ChatGPT 3.5 Turbo as the LLM server to support LLM-RAO in inferring RA solutions during the simulation.

For comparison, we implement three baseline RA schemes:

\begin{itemize}
    \item Round-Robin RA: This scheme assigns RBs sequentially to users with the highest SNRs.
    \item DRL-based RA: The state-of-the-art DRL model tailored to solve RA problem is employed \cite{Noh24}. The state is represented by users' channel and queue information, while the reward is defined as either the data rate or proportional fairness. 
    \item Genetic algorithm (GA)-based RA: This scheme can reach the globally optimal solution with infinite time \cite{coley1999}. We implement the genetic algorithm over several days to derive a naive upper bound for performance in the RA task.
\end{itemize}

\begin{table}[t!]
\centering
    \caption{Changing environment and scenario configuration with different utility functions, constraints, queuing model, and variables.
    }
    \adjustbox{width= \if 1\mycmd 0.6 \else 1.0 \fi \columnwidth}{
    \begin{tabular}{c|c|c|c|c}
    \toprule
    & Scenario 1 & Scenario 2 & Scenario 3 & Scenario 4\\ \midrule[\heavyrulewidth]\midrule[\heavyrulewidth]
    Utility & \multirow{2}{*}{Data rate} & \multirow{2}{*}{Data rate} & \multirow{2}{*}{Data rate} & Proportional \\
    function &  & & & fairness \\
    \hline
    Queue & Finite & Finite & Infinite & Finite \\ \hline
    Number of users & $K = 10 \sim 50$ & $K = 20$ & $K = 10 \sim 50$ & $K = 10 \sim 50$ \\ \hline
    Minimum rate & $R^{(k)} =0, \forall k$ & $R^{(k)} = 20 \sim 100, \forall k$ & $R^{(k)} =0, \forall k$ & $R^{(k)} =0, \forall k$ \\
    \midrule[\heavyrulewidth]\midrule[\heavyrulewidth]
    \end{tabular}
    }
    \vspace{-10pt}
    \label{tab:scenarios}
\end{table}

To evaluate the performance of LLM-RAO across different communication environments, we explore four distinct scenarios, as outlined in Table \ref{tab:scenarios}. Each scenario involves variations in the communication utility function, queuing model, number of users, and minimum rate requirements. Subsequently, we analyze its adaptability as the system transitions between scenarios and assess the problem-solving capabilities of LLM-RAO in each scenario. Note that the configured scenarios can be compared to realistic exemplary cases in a smart factory. Scenario 1 can involve varying numbers of robots efficiently handling limited tasks, while scenario 2 may prioritize urgent logistics operations, with a fixed number of robots. Scenario 3 could distribute tasks flexibly without queues to maximize production efficiency, and scenario 4 ensures fair resource allocation to maintain overall operational balance.

\subsection{Problem-Solving Capability of LLM-RAO in Optimization}

In Fig. \ref{fig:sim1234}, we plot the data rate and proportional fairness performance of LLM-RAO and baseline methods under the configured scenarios. For each scenario, the number of users, and a minimum rate, we train different DRL models, each optimized for the corresponding environment to represent the naive upper bound in terms of DRL performance. We observe that the LLM-RAO achieves up to $40\%$ performance gain compared to the DRL-based method and up to $80\%$ performance gain compared to the round-robin method. Remarkably, LLM-RAO demonstrates performance comparable to GA, showing near-optimal performance across all scenarios. Despite the inherent limitations of pre-trained LLMs in tackling mathematical optimization tasks, it is impressive to see how fine-tuning through strategic prompting effectively harnesses their advanced reasoning and inference capabilities. In contrast, due to the necessity of exploring the entire search space, i.e., all RA solutions, from scratch, the DRL-based method is prone to becoming trapped in local optima, resulting in relatively low-performance solutions.

\begin{figure*}[t]
    \centering
    \includegraphics[draft=false, width= \if 1\mycmd 1.0 \else 0.9 \fi \textwidth]{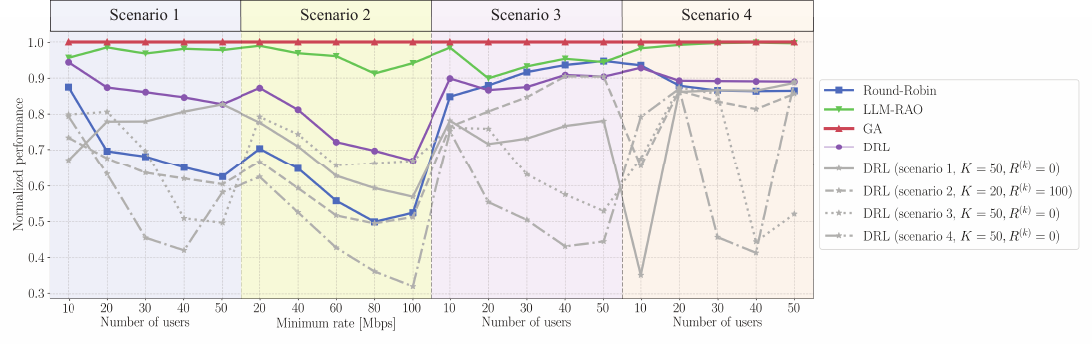}
    \vspace{-15pt}
    \caption{Evaluation of performance and adaptability of LLM-RAO and baseline methods under changing environment and scenario configuration. DRL (scenario 1, $K=50, R^{(k)} =0$) represents the DRL-based scheme using the model trained under scenario 1 with 50 users and no minimum rate constraint.}
    \label{fig:sim_all}
    \vspace{-10pt}
\end{figure*}

\subsection{Analysis of Adaptability to Environment Changes}

Fig. \ref{fig:sim_all} depicts the performance of LLM-RAO and baseline schemes in environments where the scenarios, number of users, and minimum rate requirements are continuously changing. The performance of each method is normalized by that of the naive bound (GA scheme). For the proportional fairness metric, due to its relatively small value differences, we subtracted 100 from the performance of each method before normalization.
In practical communication environments, it is infeasible to perform time-consuming retraining or to preemptively prepare models for all potential communication objectives each time the environment changes. Therefore, we evaluate the performance of DRL-based RA models trained on a single scenario, with their network weights kept completely fixed during inference. 
We observe that the LLM-RAO achieves up to $2.9$ times higher performance gain compared to the fixed DRL-based schemes. Whenever the target task and environment are switched, LLM-RAO can seamlessly adjust to the new environment via prompt modification, providing RA solutions that maximize the performance while satisfying the constraint. On the other hand, the fixed DRL models encounter difficulties in achieving high-performance solutions when the communication environment deviates from the trained scenario, as they lack the flexibility to adapt dynamically to unforeseen conditions.

\section{Conclusion}
\label{sec:conclusion}

We have proposed LLM-RAO, a novel LLM-based RA optimizer for constrained network optimization in dynamic environments. LLM-RAO offers flexibility and robustness, adapting to changes without retraining, and thereby making it highly effective for complex network challenges. In our experiments, LLM-RAO outperforms conventional DL and analytical methods, demonstrating superior problem-solving and adaptability. These results highlight its potential to enhance communication efficiency, advancing more intelligent and adaptive networks for future communication.

\bibliographystyle{IEEEtran}
\bibliography{{bibtex}}

\end{document}